\documentclass[aps,prl,twocolumn,superscriptaddress,longbibliography]{revtex4-2}

\usepackage{amssymb,amsmath,amsthm,bbm,bbold}
\usepackage{amsfonts}
\usepackage{graphicx}
\usepackage{mathtools}
\usepackage{natbib}
\usepackage[breaklinks]{hyperref}

\usepackage[normalem]{ulem}
\usepackage{color}
\usepackage{xcolor}

\def\Ho{\mathcal{H}_1}

\usepackage[normalem]{ulem}
\usepackage{color}
\usepackage{xcolor}

\newcommand{\bd}[1]{\boldsymbol{#1}}

\newcommand{\wb}{\bd{w}}

\newcommand{\F}{\mathcal{F}}

\newcommand{\E}{\mathcal{E}^N}
\newcommand{\e}{\mathcal{E}^1_N}

\newcommand{\g}{\gamma}

\newcommand{\G}{\Gamma
}

\newcommand{\Ew}{\mathcal{E}^N({\wb})}
\newcommand{\ew}{\mathcal{E}^1_N(\wb)}
\newcommand{\Tr}{\mbox{Tr}}

\newcommand{\bra}[1]{\mbox{$\langle #1 |$}}
\newcommand{\ket}[1]{\mbox{$| #1 \rangle$}}

\newcommand*\xbar[1]{\hbox{\vbox{
       \hrule height 0.6pt 
       \kern0.24ex
       \hbox{%
         \kern-0.2em
         \ensuremath{#1}%
         \kern 0.0em
         }}}}

\newcommand*\xxbar[1]{\hbox{\vbox{
       \hrule height 0.6pt 
       \kern0.3ex
       \hbox{%
         \kern-0.0em
         \ensuremath{#1}%
         \kern 0.0em
         }}}}

\newcommand{\Ebw}{\,\xxbar{\mathcal{E}}^N\!(\wb)}
\newcommand{\ebw}{\,\xxbar{\mathcal{E}}^1_N(\wb)}
\newcommand{\ebwp}{\,\xxbar{\mathcal{E}}^1_N(\wb')}

\newcommand{\ebwS}{\,\scriptsize{\xbar{\mathcal{E}}}\normalsize^1_N\hspace{-0.3mm}(\wb)}
\newcommand{\EbwS}{\,\scriptsize{\xbar{\mathcal{E}}}\normalsize^N\hspace{-0.7mm}(\wb)}

\newcommand{\Fw}{\mathcal{F}_{\!\bd{w}}}
\newcommand{\Fbw}{\xbar{\mathcal{F}}_{\!\bd{w}}}

\newtheorem{thm}{Theorem}

\begin{document}
\title{Ensemble reduced density matrix functional theory for excited states and hierarchical generalization of Pauli's exclusion principle}
\author{Christian Schilling}
\email{c.schilling@physik.uni-muenchen.de}
\affiliation{Department of Physics, Arnold Sommerfeld Center for Theoretical Physics,
Ludwig-Maximilians-Universität M\"unchen, Theresienstrasse 37, 80333 M\"unchen, Germany}
\affiliation{Munich Center for Quantum Science and Technology (MCQST), Schellingstrasse 4, 80799 M\"unchen, Germany}

\author{Stefano Pittalis}
\affiliation{CNR-Istituto Nanoscienze, Via Campi 213A, I-41125 Modena, Italy}

\begin{abstract}
We propose and work out a reduced density matrix functional theory (RDMFT) for calculating energies of eigenstates of interacting many-electron systems beyond the ground state. Various obstacles which historically have doomed such an approach to be unfeasible are overcome. First, we resort to a generalization of the Ritz variational principle to ensemble states with fixed weights. This in combination with the constrained search formalism allows us to establish a universal functional of the one-particle reduced density matrix.
Second, we employ tools from convex analysis to circumvent the too involved N-representability constraints.
Remarkably, this identifies Valone's pioneering work on RDMFT as a special case of convex relaxation and reveals that crucial information about the excitation structure is contained in the functional's domain. Third, to determine the crucial latter object, a methodology is developed which  eventually leads to a generalized exclusion principle. The corresponding linear constraints are calculated for systems of arbitrary size.
\end{abstract}

\maketitle
Developing a comprehensive understanding of excitations in many-body systems is of utmost importance from both a fundamental and technological point of view. For instance, quantum excitations intervene in crucial processes such as vision \cite{Vision}, define the properties of advanced materials \cite{OSEM} and of states of matter in general \cite{Kogar1314,Polaritons,AVMR2021} and give rise to distinctive functionalities of devices \cite{OLED1,OLED2}. Although modern  computational methodologies can determine the ground state energies of a wide range of systems relatively inexpensively and rather accurately \cite{GSSolved}, methodological innovations are called for handling excitations on an equal footing \cite{Methods}.

The workhorse of modern electronic structure calculations is the  Kohn-Sham formulation \cite{KS65} of
density-functional theory (DFT) \cite{HK64}. As far as excitations are concerned, its time-dependent extension could deal with them rigorously, at least in principle \cite{RG84}. In practice, however, the widely used time-dependent DFT is not only blessed but unfortunately also cursed by the so-called adiabatic approximation \cite{Neepa2004-Double,Maitra16,Elliot2011-Double}. Circumventing at least some of the deficiencies of adiabatic time-dependent DFT, ensemble DFT  has become in recent years a promising alternative for calculating excitations
\cite{T79,GOK2,F15B, YPBU17,GP17,SB18,GKP18,GP19,KF19,Fromager2020-DD,Loos2020-EDFA,MSFL2020,GSP2020,GKP2021} --- for example it can capture charge transfers, double excitations, and avoided/conical crossings.

From a general perspective, density functional theories are, however, not particularly well-suited for the description of strongly correlated systems. The particle density namely does not directly reflect the correlation strength, in striking contrast to the full one-particle reduced density matrix (1RDM) with fractional occupation numbers in case of strong correlations. This motivates one-particle reduced density matrix functional theory (RDMFT) \cite{Gilbert75} as a more suitable approach to strongly correlated quantum systems and explains why RDMFT has become an active field of research in recent years \cite{C00,M07,TLMH15,PG16,SKB17,S18,SS19,BTNRR19,GWK19,SBM19,C20b,C20a,G20,M20}. While the accuracy of ground state calculations compares favourably to those of DFT \cite{LM08}, no proper foundation for targeting excited states within RDMFT exists yet. For instance, a formal justification of a fully dynamical RDMFT is lacking and  the approach based on an adiabatic approximation to be exploited through linear response techniques turns out to be technically involved and numerically rather demanding \cite{GGB12,PG16}. Most remarkably, the RDMFT analogue of ensemble DFT for excited states has not even been considered yet, despite its numerous potential advantages over time-dependent functional theories.

In this letter, we propose and work out the ensemble version of RDMFT for calculating the energies of (selected) low-lying excited states. For this, we put forward a generalization of the Ritz variational principle which together with the constrained search formalism leads to the definition of a universal functional. The crucial ingredient which makes this method viable is a convex relaxation scheme. It allows us to circumvent the corresponding too intricate one-body N-representability constraints and leads instead to an easy-to-calculate generalization of Pauli's exclusion principle for mixed states.

\paragraph*{RDMFT in a nutshell $\&$ relevance of Valone's work.---}
We briefly recall ground state RDMFT. Here and in the following, we consider Hamiltonians of the form $H(h)=h+V$ on the $N$-fermion Hilbert space $\mathcal{H}_N\equiv\wedge^N[\mathcal{H}_1]$, where $h$ is a one-particle Hamiltonian and $V$ some fixed interaction (e.g.~Coulomb pair interaction). We denote the set of {\em pure} states $\G\equiv \ket{\Psi}\!\bra{\Psi}$ by $\mathcal{P}^N$ and the $d$-dimensional one-particle Hilbert space by $\mathcal{H}_1$.

Calculating the ground state energy $E(h)$ of $H(h)$ via the Ritz variational principle leads immediately to a universal functional of the 1RDM  \cite{LE79,Li83},
\begin{eqnarray}\label{Levy}
  E(h) &=& \min_{\G \in \mathcal{P}^N} \mbox{Tr}_N[(h+V)\G] \nonumber \\
 &=&  \min_{\gamma\in \mathcal{P}^1_N}\Big[\mbox{Tr}_1[h\gamma]+\F(\g)\Big]\,,
\end{eqnarray}
where
\begin{eqnarray}\label{Fp}
\F(\g) \equiv \min_{\mathcal{P}^N\ni \G \mapsto \gamma}\mbox{Tr}_N[V \G]\,.
\end{eqnarray}
Indeed, $\F$ is universal in the sense that it depends only on the fixed interaction $V$ but not on the one-particle Hamiltonian $h$. This version of RDMFT based on \eqref{Levy}, \eqref{Fp} has, however, not been practical at all. This is due to the fact that describing the functional's domain $\mathcal{P}^1_N\equiv N \mbox{Tr}_{N-1}[\mathcal{P}^N]$ of pure $N$-representable 1RDMs $\g$ has been an almost impossible task. Only recently, a formal solution to this problem has been found \cite{KL06,AK08}. Yet, the corresponding generalized Pauli constraints defining $\mathcal{P}^1_N$ could be calculated so far only for systems of up to five electrons and eleven spin-orbitals \cite{BD72,KL06,AK08,S18atoms}.

It has been Valone's crucial idea \cite{V80} to apply the constrained search formalism \eqref{Levy} by relaxing the Ritz variational principle from pure to \emph{all} ensemble states $\G \in \E$. In analogy to \eqref{Levy}, \eqref{Fp} this then leads to a universal functional
\begin{eqnarray}\label{Fe}
\xbar{\F}(\g)  &\equiv&  \min_{\E \ni \G \mapsto \g}\mbox{Tr}_N[V \G ]\,,
\end{eqnarray}
defined on the larger domain $\e \equiv  N \mbox{Tr}_{N-1}[\E]$ of ensemble $N$-representable 1RDMs. Since the latter is just described by the simple Pauli exclusion principle constraints \cite{C63}, restricting the eigenvalues $\lambda_i$ of $\g$ as $0\leq \lambda_i \leq 1$, Valone's work \cite{V80} has marked the starting point of RDMFT, at least from a practical perspective. Finally, we would like to stress that \,$\xbar{\F}$ follows as the lower convex envelope of $\F(\g)$,\, $\xbar{\F} \equiv \mbox{conv}(\F)$ \cite{S18}. As the following will show, this key result has its origin in a fruitful geometrical structure which will be pivotal to our approach.

\paragraph*{Ensemble-RDMFT for excited states.---}
To develop an RDMFT for targeting the excitation spectrum we resort to the generalization \cite{GOK88a} of the Ritz variational principle to ensemble states with fixed spectrum: let $H$ be a Hermitian operator on a $D$-dimensional Hilbert space with increasingly-ordered eigenvalues $E_j$ and eigenstates $\ket{\Psi_j}$ and denote by  $\Ew$ the set of density operators $\G$ with decreasingly-ordered spectrum $\wb\equiv (w_1,\ldots, w_D)$. Then, the following variational principle can be proven in a straightforward manner \cite{GOK88a}
\begin{equation}\label{ExVarP}
E_{\wb} \equiv \sum_{j=1}^{D} w_j E_j = \min_{\G \in \Ew} \Tr[H \G]\,,
\end{equation}
and the minimizer of the right-hand side follows as $\G_{H,\wb}=\sum_{j=1}^{D} w_j \ket{\Psi_j}\!\bra{\Psi_j}$.

At this point, it is crucial to appreciate that the knowledge of the function $E_{\wb}$ would obviously allow one to determine various excitation energies $E_j$. 
In analogy to the derivation of ensemble DFT for excited states by Gross, Oliviera and Kohn \cite{GOK88a,GOK88b}, the variational principle \eqref{ExVarP} is the key ingredient for establishing an RDMFT for excited states. Combining the constrained search \eqref{Levy} and the variational principle \eqref{ExVarP} with $H \equiv H(h)$ leads immediately to a universal functional of the 1RDM,
\begin{eqnarray}\label{Fw}
\Fw(\g) &\equiv& \min_{\Ew\ni\Gamma\mapsto \g}\mbox{Tr}_N[V\Gamma]\,.
\end{eqnarray}
In practice, one would restrict this $\wb$-RDMFT to just a few finite weights $w_1,\ldots,w_r$ and the minimization of the total energy functional $\mbox{Tr}_1[h \g]+ \Fw(\g)$ would eventually yield the energy $E_{\wb}$.
It is also worth stressing, that pure \emph{ground state} RDMFT is included in our general $\wb$-RDMFT as the special case $\wb_0 \equiv (1,0,0,\ldots)$. This observation also implies, however, that finding a practically useful description of the underlying domain $\ew$ of \emph{$\wb$-ensemble} $N$-representable 1RDMs is impossible, at least for realistic system sizes \cite{KL04,KL06,AK08}.

This fundamental concern also explains why ensemble RDMFT for calculating excitation energies has never been established.
It will therefore be a major achievement of our work to find and work out in the following a methodology for circumventing the too intricate one-body $\wb$-ensemble $N$-representability constraints.

\paragraph*{Convex relaxation.---}
\begin{figure}[htb]
\includegraphics[scale=0.57]{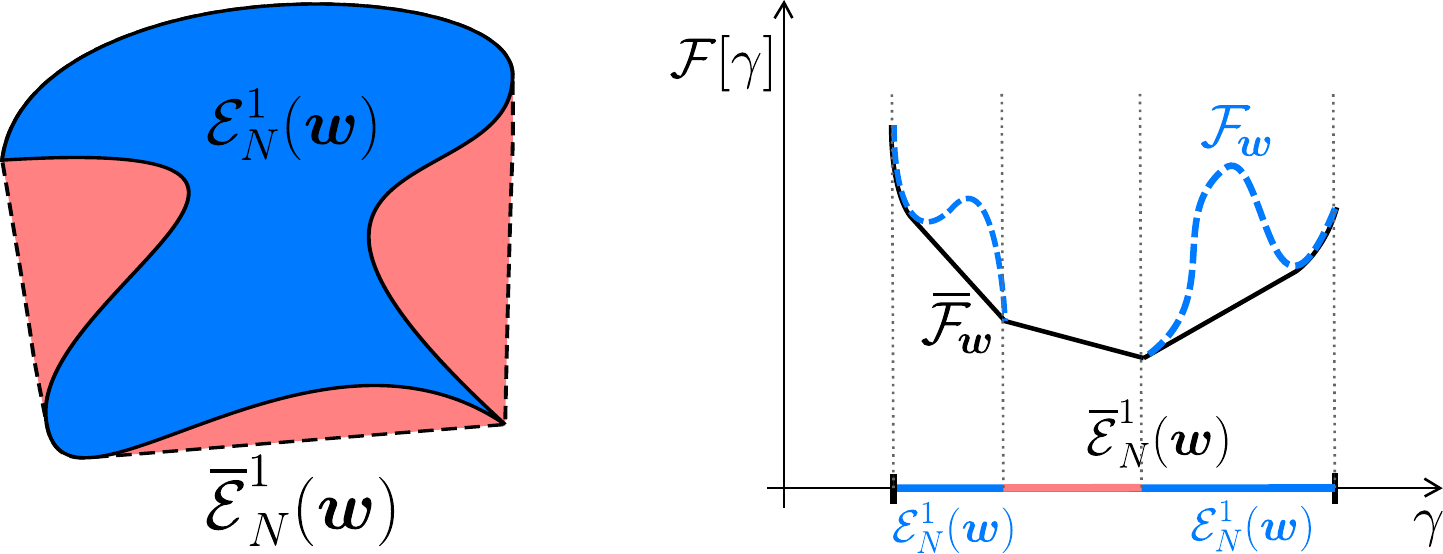}
\caption{Schematic illustration of the convex hull $\ebw$ of the ``blue'' set $\ew$ (left) and the convex envelope $\,\Fbw$ of $\Fw$ (right). See text for more details.}
\label{fig:relax}
\end{figure}
Given the prominence of the Ritz variational principle in quantum physics, Valone's idea to apply the constrained search formalism to the larger set $\E$ of ensemble states was rather natural. Since the generalization of this seminal idea to the variational principle \eqref{ExVarP} is less obvious, we take the relation \,$\xbar{\F} \equiv \mbox{conv}(\F)$ \cite{S18} in ground state RDMFT as an inspiration: A prominent concept in convex analysis explains that any minimization problem, at least in principle, can be turned into a convex one without altering the result. This exact convex relaxation applied in our context is illustrated in Figure \ref{fig:relax}: First, one extends the ``blue'' domain of $\Fw$ to its convex hull $\ebw = \mbox{conv}(\ew)$ by adding the ``red'' region and defining there $\Fw\equiv \infty$. Then, one replaces $\Fw$ by its lower convex envelope
\begin{equation}\label{Forel}
\Fbw \equiv \mbox{conv}\left(\Fw\right)\quad \mbox{on}\quad \ebw \equiv \mbox{conv}(\ew)\,.
\end{equation}
Per construction, the sought-after energy $E_{\wb}$ can now be obtained by minimizing the energy functional $\mbox{Tr}_1[h \g]+\Fbw(\g)$ on the set $\ebw$,
\begin{equation}\label{EwFbw}
E_{\wb} = \min_{\g \in \ebwS} \left[\mbox{Tr}_1[h \g]+\Fbw(\g) \right]\,,
\end{equation}
rather than $\mbox{Tr}_1[h \g]+\Fw(\g)$ on $\ew$. This convex relaxation of $\wb$-ensemble RDMFT has two pleasant and far-reaching consequences. As we will show below, a compact description of the convex set $\ebw$ can be found \emph{and} the convexity of \,$\Fbw$ implies that the  minimization cannot get stuck in local minima.

We conclude this section by presenting an equivalent but constructive expression for \,$\Fbw$.
As it is shown in the supporting information \footnote{See Supplemental Material at url for a proof
of relations \eqref{FwLevy}, \eqref{Ebwmajor} and an illustration of the polytopes $\Sigma(\wb)$.}, the definition of the lower convex envelop leads in a straightforward (but technical) manner to
\begin{eqnarray}\label{FwLevy}
\Fbw(\g) \equiv  \min_{\EbwS\ni\G\mapsto \g}\mbox{Tr}_N[V\G]\,.
\end{eqnarray}
The underlying search space $\Ebw$ is nothing else than the convex hull of $\Ew$, which can also be characterized as \cite{Note1}
\begin{equation}\label{Ebwmajor}
\Ebw  \equiv  \mbox{conv}(\Ew)= \{\G \in \E| \mathrm{spec}(\G) \prec \wb\}\,.
\end{equation}
Here, $\bd{v}$ is said to be majorized by $\wb$, $\bd{v}\prec\,\wb$, if for all $k=1,2,\ldots,D$ one has
\begin{equation}\label{majorize}
v^\downarrow_1+\ldots + v^\downarrow_k \leq w^\downarrow_1+\ldots +w^\downarrow_k\,,
\end{equation}
where $\bd{v}^\downarrow, \bd{w}^\downarrow$ denote the vectors with the same entries, but sorted in descending order.
Intriguingly, relations \eqref{FwLevy} and \eqref{Ebwmajor} reveal that the functional \,$\Fbw$ could also have been defined in the spirit of Valone's work \cite{V80} by replacing in \eqref{ExVarP} the set $\Ew$ by its convex hull $\Ebw$. This modification of \eqref{ExVarP} can be seen as the historically missed variational principle for establishing a \emph{viable} ensemble RDMFT for excited states. Moreover,
since the partial trace $\mbox{Tr}_{N-1}[\cdot]$ is linear we obtain
\begin{equation}\label{ebw}
\ebw = N \mbox{Tr}_{N-1}[\Ebw]\,.
\end{equation}

\paragraph*{Calculation of functional domain $\ebw$.---}
In general, without knowing the functional domain, the common process of developing more and more accurate and sophisticated approximations to the universal functional cannot be initiated. Also, having just a formal definition of $\ebw$ as in Eq.~\eqref{Forel} or \eqref{ebw} is actually not sufficient. Instead, a concrete description is needed, allowing one in minimization algorithms to easily check whether a given 1RDM $\g$ belongs to $\ebw$. In the following, we achieve the ideal scenario: A description of $\ebw$ in terms of just a few linear inequalities is found, similar to the Pauli exclusion principle constraints in ground state RDMFT.

\begin{figure}[htb]
\includegraphics[scale=0.49]{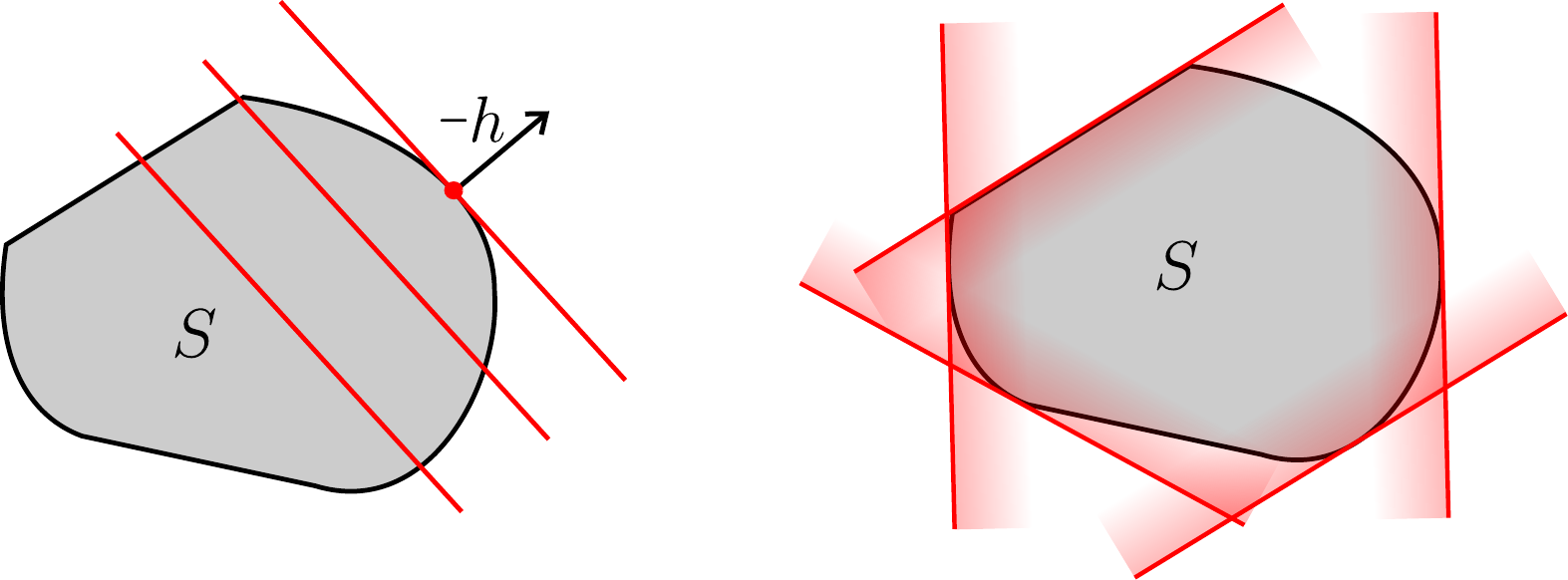}
\caption{Geometric illustration of the dual characterization of a convex, compact set $S$, based on the minimization of all possible linear functions (see text for more details).}
\label{fig:dual}
\end{figure}

For this we resort to analytical tools some of which have extensively been used in quantum chemistry since the
1960s (see, e.g., \cite{C63,CY00} for a comprehensive introduction). The first one is a very well-known duality correspondence
which is illustrated in Figure \ref{fig:dual} for a general convex, compact subset $S$ of an Euclidean space: the minimization of a linear function $\langle \cdot, h \rangle$ on $S$ means to shift the hyperplane of constant value $\langle \g, h\rangle$ (shown as ``red'' lines in the left panel) along its normal direction $-h$ until the boundary is reached. By realizing such minimizations for all possible `directions' $-h$, we obtain a complete characterization of $S$ through its boundary points.

In our context, dual characterization of $\ebw$ means to study the minimization of $\mbox{Tr}_1[h \g]$
on the convex, compact set $\ebw$ for all Hermitian operators $h$ on the one-particle Hilbert space $\Ho$.
It is exactly this aspect which reveals a fruitful equivalence of our theoretical problem of characterizing the set $\ebw$
and describing systems of $N$ non-interacting fermions with a one-particle Hamiltonian $h$.
Since $\ebw$ is invariant under unitary conjugation, $u\, \ebw\, u^\dagger = \ebw$, we can restrict to $h$
with a fixed eigenbasis, $h=\sum_{i=1}^{d}h_i \ket{i}\!\bra{i}$, and increasingly-ordered energies $h_j$.
To proceed, since the definition \eqref{ebw} of $\ebw$ refers to the set $\Ebw$, we lift our minimization problem from the one- to the  $N$-particle level, according to
 \begin{equation}\label{Min1liftN}
    \min_{\g \in \ebwS} \mbox{Tr}_1[h \g] = \min_{\G \in \EbwS} \mbox{Tr}_N[h \G]\,.
    \end{equation}
The minimizers of the right side of \eqref{Min1liftN} then lead via
\begin{equation}\label{sequence}
h \mapsto \G_{h,\wb} \mapsto \g_{h,\wb} \mapsto  \bd{\lambda}_{h,\wb} 
\end{equation}
to all extremal points $\bd{\lambda}_{h,\wb}$ of the polytope
\begin{equation}\label{Sigma}
\Sigma(\wb) \equiv \mbox{spec}\big(\ebw\big)
\end{equation}
of natural occupation numbers $\bd{\lambda}\equiv \mbox{spec}(\g)$.
Just to recall, determining $\Sigma(\wb)$ is sufficient for the description of $\ebw$ because of its unitary invariance.

To determine the required minimizers $\G_{h,\wb}$ of the right-hand side of \eqref{Min1liftN}, notice that the eigenstates of the one-particle Hamiltonian $h$ are given by the configuration states $\bd{i}\equiv \ket{i_1,\ldots,i_N}$ with energies $\sum_{j=1}^N h_{i_j}$. Consequently,
as explained by the variational principle \eqref{ExVarP}, the minimizers follow as
\begin{equation}\label{GOK0int}
\G_{h,\wb} = \sum_{j=1}^{D} w_j \ket{\bd{i}_j}\!\bra{\bd{i}_j}\,,
\end{equation}
where $\bd{i}_j$ is the $N$-fermion configuration with the $j$-th lowest energy. Due to the noninteracting character, one can easily determine for any $\G_{h,\wb}$ its natural occupation numbers $\bd{\lambda}$ as required by \eqref{sequence}.

To obtain the \emph{vertex representation} of the polytope $\Sigma(\wb)$, it remains to determine for each choice $h_1 \leq \ldots \leq h_d$ the corresponding sequence of eigenstates $\ket{\bd{i}_j}$ ordered according to their energy. Since there are only finitely many different sequences, this amounts to a purely combinatorial problem.
The example of $N=3$ fermions with $r=3$ finite weights, $\wb=(w_1,w_2,w_3,0,\ldots)$, will be sufficiently representative for the general case. According to \eqref{GOK0int}, we need to determine all possible sequences $\bd{i}_1, \bd{i}_2, \bd{i}_3$ of the three energetically lowest configurations. Independent of the values of various $h_j$, the first two configurations are always given by $\bd{i}_1=(1,2,3)$ and $\bd{i}_2=(1,2,4)$. The third lowest will be either $(1,2,5)$ or $(1,3,4)$, depending on the ordering between $h_2+h_5$ and $h_3+h_4$.
Consequently, there are in total two different minimizers $\G_{h,\wb}$ and according to \eqref{sequence} two vertices of decreasingly ordered natural occupation numbers,
\begin{eqnarray}\label{vertices}
\bd{v}^{(1)}&=&(1,1,w_1,w_2,w_3,0,\ldots) \nonumber \\
\bd{v}^{(2)}&=&(1,w_1+w_2,w_1+w_3,w_2+w_3,0,\ldots)\,.
\end{eqnarray}
The polytope $\Sigma(\wb)$ is eventually obtained as the convex hull of $\bd{v}^{(1)}, \bd{v}^{(2)}$ and all their permutations. The supporting information \cite{Note1} provides a graphical illustration of $\Sigma(\wb)$ for different $\wb$.

\begin{figure}[htb]
\frame{\includegraphics[scale=0.52]{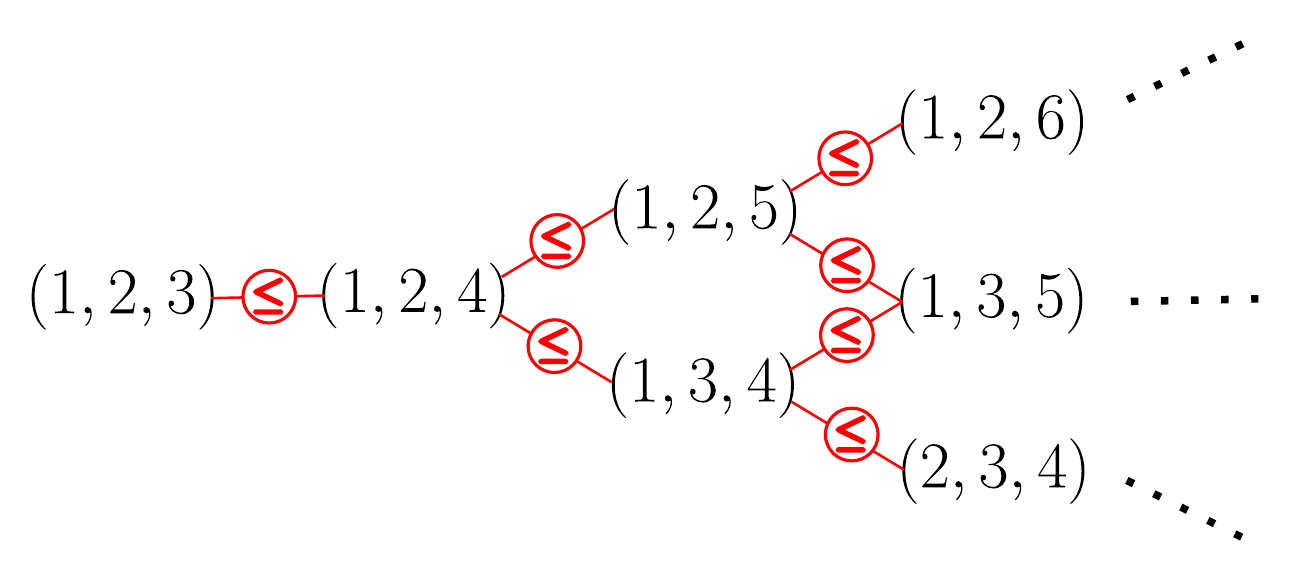}}
\caption{Illustration of the excitation spectrum for the case of $N=3$ fermions (see text for more details).}
\label{fig:excit}
\end{figure}
As a help for dealing with larger numbers $r$ of nonvanishing weights, we illustrate in Figure \ref{fig:excit} the so-called Gale order \cite{bookGale}. All configurations are systematically arranged, where the $\leq$-sign between two configurations $\bd{i}$ and $\bd{j}$ means that $\bd{i}$ has always a lower energy than $\bd{j}$ for all $h$. Figure \ref{fig:excit} also confirms
that the first two configurations are always given by $\bd{i}_1=(1,2,3)$, $\bd{i}_2=(1,2,4)$ and that the dimension $d$ of the one-particle Hilbert space does not play any role. In addition, it shows that increasing the particle number from $3$ to $N$ will not change the excitation structure but just add another $N-3$ `1's at the very beginning of each spectral vector in \eqref{vertices}.

\paragraph*{Generalization of Pauli's exclusion principle.---}
The vertex representation of the polytope $\Sigma(\wb)$ can also be turned into a \emph{halfspace representation}, leading according to \eqref{Sigma} to the desired practical description of the domain $\ebw$. This mathematical procedure for \emph{arbitrary} $r$ will be discussed in \cite{CLLS21} and we present it here only for $r=1,2$. These two cases are particularly relevant from a physical point of view since the corresponding $\wb$-RDMFT describes the ground state energy and its gap.

For $r=1$ and $r=2$, the permutation-invariant polytope $\Sigma(\wb)$ is generated by only one vertex, $(1,\ldots,1,0,\ldots)$ and $(1,\ldots,1,w_1,w_2,0,\ldots)$, respectively. According to Rado's theorem \cite{R52}, a vector $\bd{\lambda}$ lies in such distinctive polytope if and only if $\bd{\lambda}$ is majorized by the vertex $\bd{v}^{(1)}$, $\bd{\lambda}\prec\bd{v}^{(1)}$. The definition \eqref{majorize} of majorization is nothing else than the halfspace representation of $\Sigma(\wb)$.

Most remarkably, referring to different values $r$, there is a \emph{hierarchy} of linear inequalities  for $\Sigma(\wb)$ which generalize Pauli's exclusion principle. On its lowest level, $r=1$, one recovers the Pauli exclusion principle $\lambda^{\downarrow}_1 \leq 1$. For $r=2$, one additional constraint,
\begin{equation}\label{GPCr2}
\sum_{j=1}^{N}\lambda^{\downarrow}_j \leq N-1+w_1\,,
\end{equation}
occurs. This already manifests a generalization of the exclusion principle. Next, for $r=3$ again one additional constraint emerges, namely
\begin{equation}\label{GPCr3}
2\sum_{j=1}^{N-1}\lambda^{\downarrow}_j+\lambda^{\downarrow}_N+\lambda^{\downarrow}_{N+1} \leq 2N-2+w_1+w_2\,.
\end{equation}
\begin{table}[htb]
\setlength{\tabcolsep}{5pt} 
\renewcommand{\arraystretch}{1.1}
\begin{tabular}{|c|c|c|c|c|c|c|c|c|c|}
\hline
r       & 1& 2& 3& 4&  5&  6&  7&   8& 9  \\ \hline
$\#\bd{v}^{(k)}$    & 1& 1& 2& 4& 10& 28& 90& 312& 1160  \\ \hline
$\#ineq$ & 1& 2& 3& 4&  8& 13& 23&  42&   88 \\ \hline
\end{tabular}
\caption{Number $\#$ of generating vertices $\bd{v}^{(k)}$ and inequalities of permutation-invariant polytope $\Sigma(\wb)$}
\label{tab:r}
\end{table}
The inequalities of this hierarchy for larger $r$ are presented in \cite{CLLS21}, while in Table \ref{tab:r} we just list the number of generating vertices $\bd{v}^{(k)}$ and inequalities of $\Sigma(\wb)$ for $r \leq 9$.
It is worth recalling that all these results are independent of the particle number $N$ and the dimension $d$ of the one-particle Hilbert space (provided $N \geq r-1$, $d \geq N+r-1$).

\paragraph*{Outlook.---}
We have proposed and worked out in the form of $\wb$-RDMFT a viable generalization of ground state RDMFT for targeting the energies of the first few excitations. The crucial ingredient has been a convex relaxation scheme which allowed us to overcome the obstacles which have historically doomed such an approach to be unfeasible.
As a major achievement, our work has namely overcome the too involved one-body $\wb$-ensemble $N$-representability constraints. Instead, a hierarchy of easy-to-calculate generalizations of Pauli's exclusion principle constraints has been revealed, providing a practical description of the functional domain $\ebw$.

We expect a broad significance of those novel constraints across the quantum sciences.
For instance, since they describe the compatibility of $N$-fermion and one-fermion density operators our work solves a certain class of quantum marginal problems. The latter play a crucial role, e.g., for quantum communication and quantum information processing \cite{KL04,DH05}. Moreover, since realistic systems in nature are described by \emph{mixed} states due to the \emph{finite} temperature, it is primarily not Pauli's exclusion principle which dictates their physics but our generalized exclusion principle for mixed states. The spectral polytopes $\Sigma(\wb)$ are also strongly related to a possible generalization of the Fermi-Dirac distribution to \emph{interacting} fermions: Increasing the temperature $T$ of a system makes the spectrum $\wb(T)$ of the Gibbs state $\G(T)\propto e^{-H/k_B T}$ more mixed which in turn reduces the size of the polytope $\Sigma(\wb(T))$. In that sense the generalized exclusion principle constraints provide a tool to determine the maximal temperature of a system which is still compatible to given occupation numbers.

Equipped with the definition \eqref{FwLevy} of the universal functional \,$\Fbw$ and the practical description of its domain $\ebw$,
the common process of developing more and more accurate and sophisticated functional approximations can be initiated. Actually, the ground state functional may serve as a first approximation to $\,\Fbw$, as it is explained in the Supporting Information \cite{Note1}: At least for weakly interacting systems, the energy difference $E_{\wb} \leq E_{\wb'}$ (where $\wb' \prec \wb$) is primarily a direct \emph{geometrical} consequence, following from $\ebwp \subset \ebw$. In particular, knowing the boundary of $\ebw$ is sufficient for determining approximately $E_{\wb}$. Another promising and particularly sophisticated strategy would be to work out in the context of mixed states exactly the same three steps \cite{M84,BB02,GPB05} that led to the rather accurate BBC ground state functionals.

\begin{acknowledgments}
We thank F.\hspace{0.5mm}Castillo, J.-P.\hspace{0.5mm}Labb\'e and J.\hspace{0.5mm}Liebert for helpful discussions. We acknowledge financial support from the Deutsche Forschungsgemeinschaft (Grant SCHI 1476/1-1) and the UK Engineering and Physical Sciences Research Council (Grant EP/P007155/1) (C.S.) and from the MIUR PRIN Grant No. 2017RKWTMY (S.P.).
\end{acknowledgments}

\bibliography{Refs3}

\onecolumngrid
\newpage
\begin{center}\Large{\textbf{Supplemental Material}}
\end{center}
\setcounter{equation}{0}
\setcounter{figure}{0}
\setcounter{table}{0}
\makeatletter
\renewcommand{\theequation}{S\arabic{equation}}
\renewcommand{\thefigure}{S\arabic{figure}}
\vspace{0.5cm}

\section{Proof of relation (8) and (9)}
We recall relation (8) and (9) as part of the following theorem.
\begin{thm}\label{Fbwmajor}
The relaxed $\wb$-ensemble functional $\,\Fbw$ follows as
\begin{equation}\label{Fbwmajor}
\Fbw(\g) = \min_{\EbwS \ni \G \mapsto \g} \mathrm{Tr}[\G V]\,,
\end{equation}
and in addition we have
\begin{eqnarray}
\Ebw  &=& \!\bigcup_{\wb'\prec\,\wb}\!\mathcal{E}^N(\wb')\equiv \{\G \in \E| \mathrm{spec}(\G) \prec \wb\} \label{Ebwmajor}  \\
\ebw &=& N\Tr_{N-1}[\Ebw]= \bigcup_{\wb'\prec\,\wb} \mathcal{E}^1_N(\wb')\,.\label{ebwmajor}
\end{eqnarray}
\end{thm}
Here (as in the paper), $\Ebw \equiv \mbox{conv}(\Ew)$ is defined as the convex hull of the (non-convex) set $\Ew$ of $N$-fermion density operators $\G$ with spectrum $\wb$.
Moreover, $\bd{v}$ is said to be majorized by $\wb$, $\bd{v}\prec\,\wb$, if for all $k=1,2,\ldots,D$ one has
\begin{equation}\label{majorize}
v^\downarrow_1+\ldots + v^\downarrow_k \leq w^\downarrow_1+\ldots +w^\downarrow_k\,,
\end{equation}
where $\bd{v}^\downarrow, \bd{w}^\downarrow$ denote the vectors with the same entries, but sorted in descending order.

\begin{proof}
Our proof of Theorem 1 makes use of the following well-known theorem \cite{AU82}
\begin{thm}[Uhlmann]\label{thm:Uhl}
Let $\G$ and $\G'$ be two density operators on a $D$-dimensional complex Hilbert space.
Then there exist unitary operators $U_i$ and weights $0\leq p_i\leq 1$, $\sum_{i}p_i=1$, such that
\begin{equation}
\G' = \sum_{i=1}^D p_i U_i \G U_i^\dagger
\end{equation}
if and only if $\mbox{spec}(\G') \prec \mbox{spec}(\G)$.
\end{thm}

Equation \eqref{Ebwmajor} follows directly from Uhlmann's theorem. To explain this, we first observe that the set $\Ew$ can be parameterized as the family of all unitary conjugations of some arbitrary $\G \in \Ew$, $\Ew = \{U\G U^\dagger\}$. Indeed, for any $\G' \in \Ew$ the unitary transformation $U$ then just maps the eigenstates of $\G$ to those of $\G'$.
Accordingly, Theorem \ref{thm:Uhl} states that a density operator $\G'\in \E$ can be written as a convex combinations of $\G_i \in \Ew$ (actually $\G_i \equiv U_i \G U_i^\dagger$) if and only if its spectrum is majorized by $\wb$. Relation \eqref{ebwmajor} follows then directly from the linearity of the partial trace and Eq.~\eqref{Ebwmajor}. To prove \eqref{Fbwmajor}, we use the fact that each element $\G \in \Ebw \equiv \mbox{conv}(\Ew)$ can be written as a convex combination $\sum_i p_i \G_i$ of $\G_i \in \Ew$, leading to
\begin{eqnarray}
\lefteqn{\min_{\EbwS \ni \G \mapsto \g} \mathrm{Tr}[\G V]} \hspace{1cm} \nonumber  \\
&=&\min_{\scriptsize\begin{array}{c}\sum_i p_i\G_i \mapsto \g, \\ \G_i \in \Ew \end{array}} \sum_i p_i \mathrm{Tr}[\G_i V] \nonumber  \\
&=& \min_{\scriptsize\begin{array}{c}\sum_i p_i\g_i=\g, \\ \g_i \in \ew \end{array}} \min_{\{\Ew \ni \G_i \mapsto \g_i\}}\sum_i p_i \mathrm{Tr}[\G_i V] \nonumber \\
&\equiv&  \min_{\scriptsize\begin{array}{c}\sum_i p_i\g_i=\g, \\ \g_i \in \ew \end{array}} \sum_i p_i \Fw(\g_i) = \Fbw(\g)\,.
\end{eqnarray}
\end{proof}

\section{Spectral polytopes and role of their boundary}
We consider the example of $N=2$ fermions with a $d=3$ dimensional one-particle Hilbert space. We choose those small values for $N,d$ since this will allow us to graphically illustrate the spectral polytopes $\Sigma(\wb)$. At the same time, this setting already exhibits all relevant features of arbitrary settings $N,d$, including the complete basis set limit,  $d \rightarrow \infty$.
To target the lowest three eigenstates of a given Hamiltonian, we consider three exemplary weight vectors:
\begin{equation}\label{wexamples}
\wb^{(A)}=(1,0,0)\,, \quad \wb^{(B)}=(0.7,0.3,0)\,,\quad  \wb^{(C)}=(0.5,0.3,0.2)\,.
\end{equation}
For a given class
\begin{equation}\label{ham}
H(h) \equiv h+W
\end{equation}
of Hamiltonians characterized by a fixed pair interaction $W$ each choice of $\wb$ leads to a corresponding universal functional
$\Fbw$  on a specific domain $\ebw$ described by the polytope $\Sigma(\wb)$. The spectral polytopes $\Sigma(\wb)$ for the three exemplary cases \eqref{wexamples} are shown in Fig.~1. There, we can restrict $\bd{\lambda}$ to just two entries $\lambda_1,\lambda_2$, while $\lambda_3$ follows from the normalization, $\lambda_1+\lambda_2+\lambda_3=2$. It is crucial to recall that the polytopes $\Sigma(\wb)$ depend only on $N,d$ and $\wb$ but not on any Hamiltonian. \\
\\
The first RDMFT (with $\wb^{(A)}$) is characterized by $r=1$ non-vanishing weights and is therefore nothing else than
ground state RDMFT. The choices $\wb^{(B)}$ and $\wb^{(C)}$ represent an RDMFT for targeting the lowest two and the lowest three eigenstates, respectively.

\begin{figure}[htb]
\includegraphics[scale=0.37]{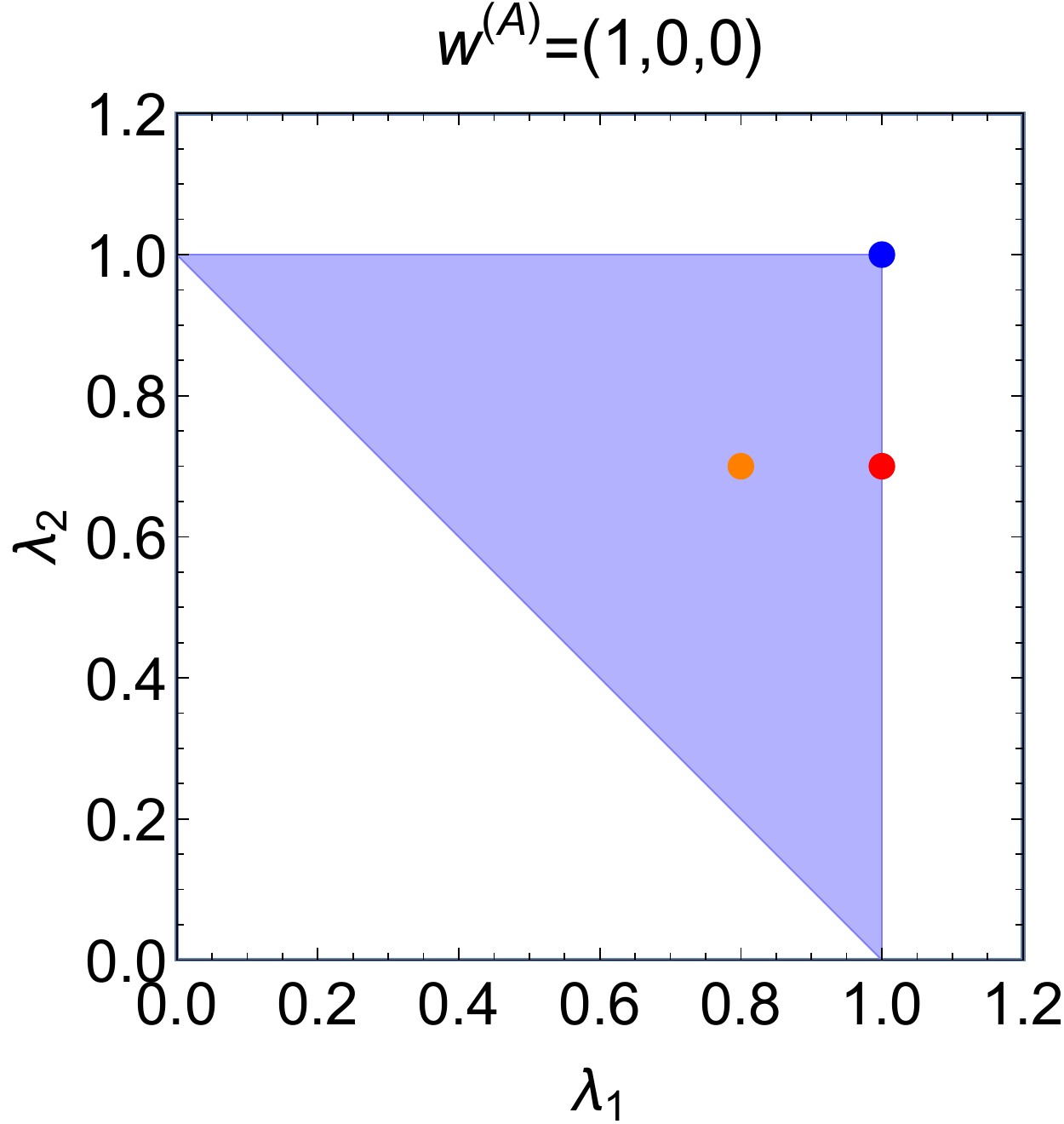}
\includegraphics[scale=0.37]{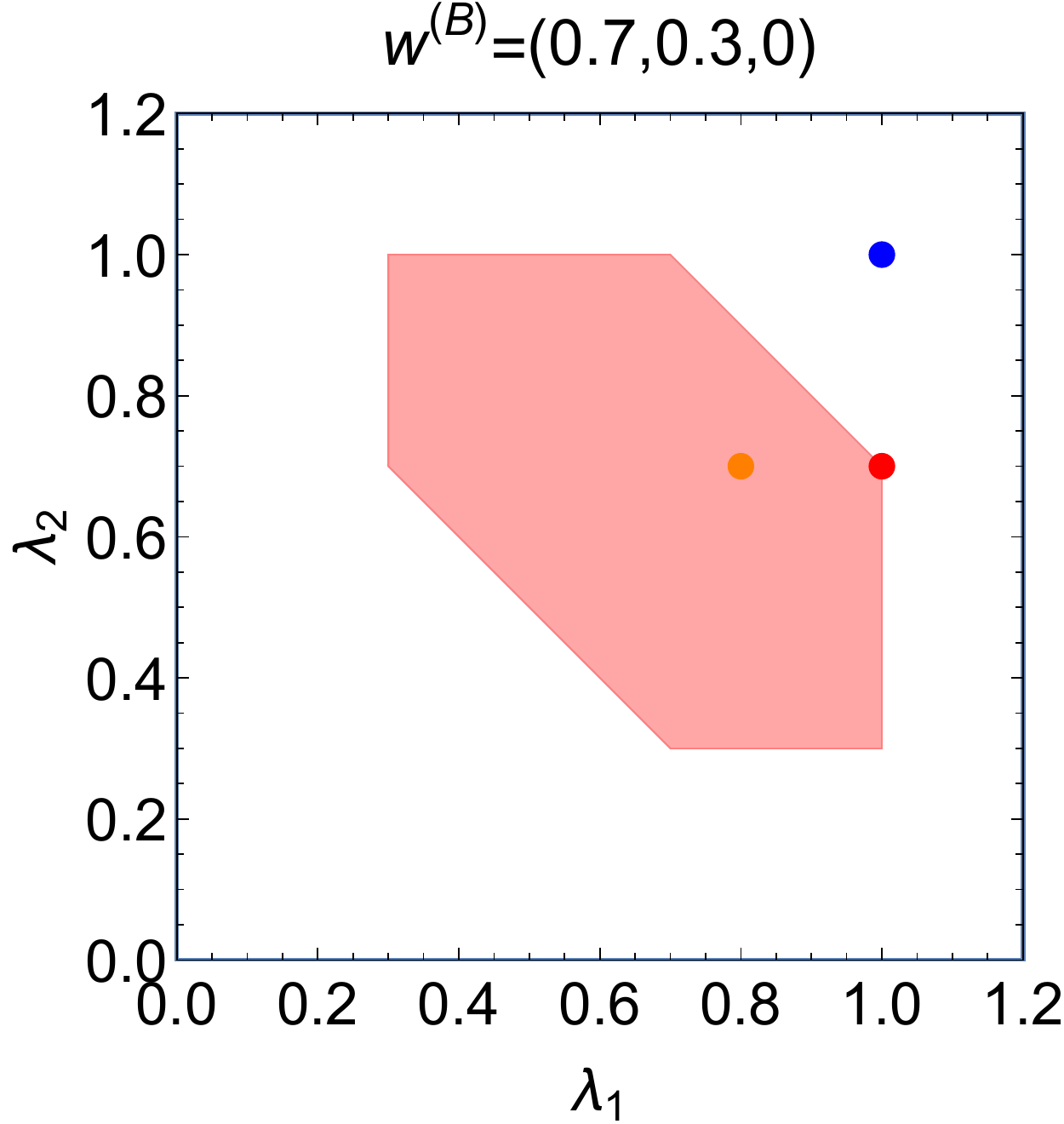}
\includegraphics[scale=0.37]{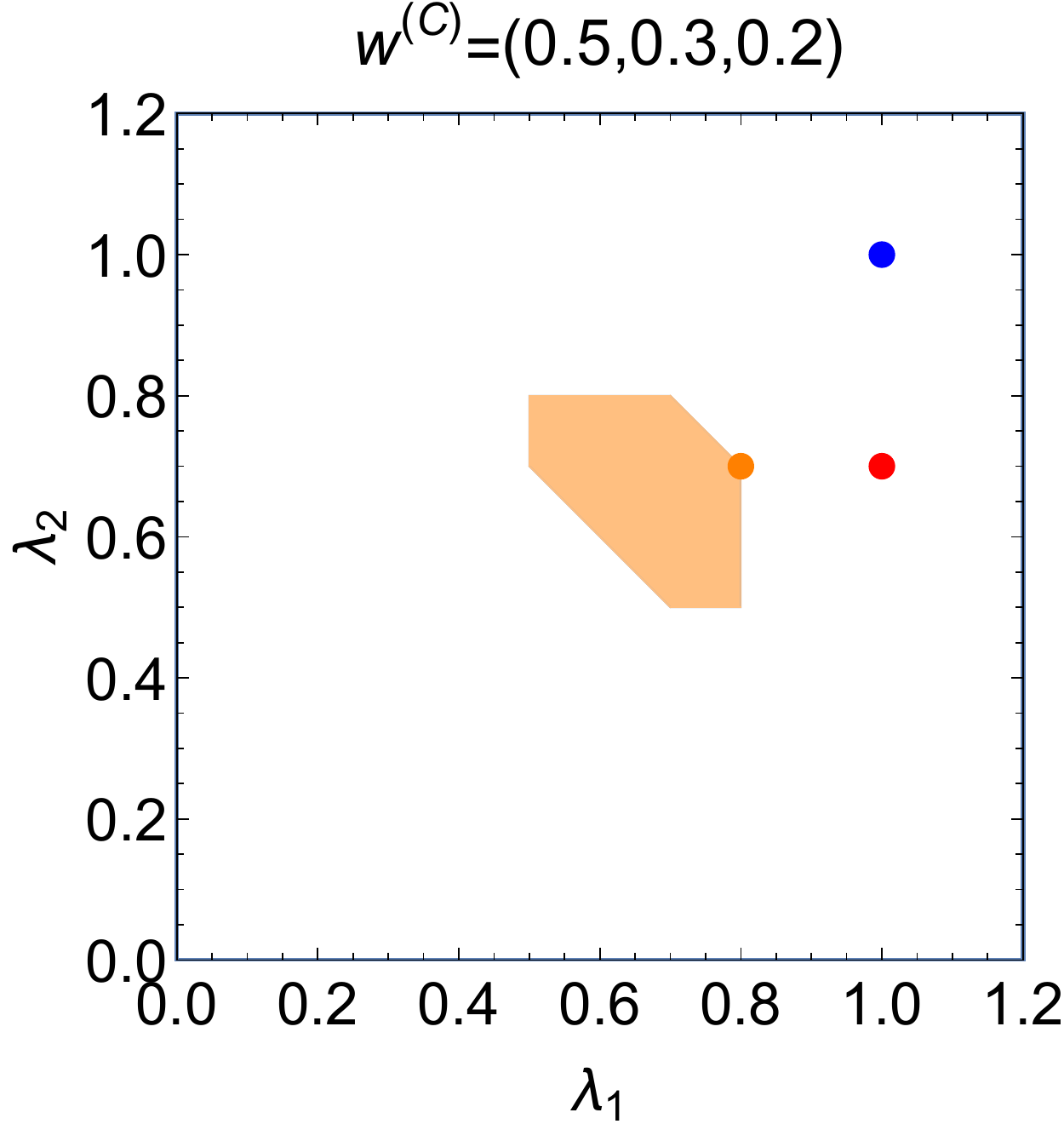}
\caption{Polytope $\Sigma(\wb)$ of possible natural occupation number vectors $\bd{\lambda}$ for $N=2$ fermions, dimension $d=3$ and the three exemplary weight vectors $\wb^{(A)}=(1,0,0)$, $\wb^{(B)}=(0.7,0.3,0)$ and $\wb^{(A)}=(0.5,0.3,0.2)$. In addition, we present the specific $\bd{\lambda}$ of the three minimizers \eqref{Gminim} for non-interacting fermions as blue, red and orange points. See text for more details.}
\label{fig:relax}
\end{figure}

Just to recall, the universal functional $\Fbw$ allows one to determine the averaged energy $w_1 E_1(h)+w_2 E_2(h)+w_3 E_3(h)$
by the following minimization
\begin{equation}\label{Emin}
E_{\wb}(h) \equiv w_1 E_1(h)+w_2 E_2(h)+w_3 E_3(h) = \min_{\g \in \ebw} \left[\mbox{tr}_1[h \g] + \Fbw(\g) \right]\,,
\end{equation}
where $E_j(h)$ denotes the $j$-th lowest eigenenergy of the Hamiltonian \eqref{ham}. This statement is true for any choice of decreasingly ordered weight vectors $\wb$, in particular also for our three examples in Eq.~\eqref{wexamples}. Hence, $\mbox{tr}_1[h \g] + \xbar{\F}_{\wb^{(A)}}(\gamma)$ needs to be minimized over the `blue' spectral polytope $\Sigma(\wb^{(A)})$ to obtain the energy $E_1(h)$,  $\mbox{tr}_1[h \g] + \xbar{\F}_{\wb^{(B)}}(\gamma)$ needs to be minimized over the `red' spectral polytope $\Sigma(\wb^{(B)})$ to obtain the energy $0.7 E_1(h)+0.3 E_2(h)$ and  $\mbox{tr}_1[h \g] + \xbar{\F}_{\wb^{(C)}}(\gamma)$ needs to be minimized over the `orange' spectral polytope $\Sigma(\wb^{(C)})$ to obtain the energy $0.5 E_1(h)+0.3 E_2(h)+0.2 E_3(h)$. From those three energy averages one can extract in particular the individual energies $E_2(h)$ and $E_3(h)$ through appropriate linear combinations.

To explain now why the boundary of the spectral polytopes $\Sigma(\wb)$ contains crucial information about the excitation spectrum we consider the case of \emph{non-interacting} fermions. This means to consider the family of Hamiltonians $\tilde{H}(h)=h$, i.e, we chose $W\equiv 0$ in \eqref{ham}. Of course, this is a rather special case since the corresponding interaction functional $\Fbw$ vanishes, $\Fbw \equiv 0$. Nonetheless, one can still calculate in the same manner as described above the eigenenergies $\tilde{E}_{\wb}(h)$ for the Hamiltonians  $\tilde{H}(h)\equiv h$.
We added a `tilde' to distinguish the case of non-interacting fermions from the one of interacting fermions. In the same manner as above, we could obtain the energy $\tilde{E}_{\wb}(h)$ as
\begin{equation}
\tilde{E}_{\wb}(h) =\min_{\g \in \ebw} \mbox{tr}_1[h \g]\,,
\end{equation}
where we already used the fact that the interaction functional is zero. For our three examples in Eq.~\eqref{wexamples}, we would then minimize the rather trivial energy functional $\mbox{tr}_1[h \g]$ over the three different polytopes shown in Fig.~1 to obtain the corresponding weighted energies
$\tilde{E}_1(h)$, $0.7 \tilde{E}_1(h)+0.3 \tilde{E}_2(h)$ and $0.5 \tilde{E}_1(h)+0.3 \tilde{E}_2(h)+0.2 \tilde{E}_3(h)$, respectively. Since the energy functional is linear we can even determine the minimizers $\bd{\lambda}$ for those three cases and also the corresponding $2$-fermion density operators:
\begin{eqnarray}\label{Gminim}
\G^{(A)}&=&\ket{1,2}\!\bra{1,2} \hspace{5.57cm}\mapsto  \quad \color{blue}{\bd{\lambda}^{(A)}=(1,1,0)}\\
\G^{(B)}&=&0.7\, \ket{1,2}\!\bra{1,2}+0.3\, \ket{1,3}\!\bra{1,3}\hspace{2.705cm}\mapsto \quad \color{red}{\bd{\lambda}^{(B)}= (1,0.7,0.3)}\nonumber \\
\G^{(C)}&=&0.5\, \ket{1,2}\!\bra{1,2}+0.3\, \ket{1,3}\!\bra{1,3}+0.2\, \ket{2,3}\!\bra{2,3}\quad\mapsto \quad \color{orange}{\bd{\lambda}^{(C)}= (0.8,0.7,0.5)} \nonumber
\end{eqnarray}
The three minimizers $\bd{\lambda}^{(A)}, \bd{\lambda}^{(B)}, \bd{\lambda}^{(C)}$ are presented in Fig.~\ref{fig:relax}. Since the functional $\mbox{tr}_1[h \g] \equiv h_1 \lambda_1+h_2 \lambda_2+h_3 \lambda_3$ ($h_i$ denote the eigenvalues of $h$) is linear, they lie on the boundary of their respective polytopes. To be more specific, for any $\wb$ the minimizer is give by
that vertex $\bd{v}\equiv (v_1,v_2,v_3)$ of $\Sigma(\wb)$ which minimizes the energy $h_1 v_1 + h_2 v_2 + h_3 v_3$.\\
\\
The crucial observation is the following: the minimizer $\bd{\lambda}^{(A)}$ of $\mbox{tr}_1[h \g]$ on the `blue' polytope corresponding to $\wb^{(A)}$ cannot be the minimizer of $\mbox{tr}_1[h \g]$ on the `red' polytope corresponding to $\wb^{(B)}$.  This is because $\bd{\lambda}^{(A)}$ lies \emph{outside} of the `red' polytope. Consequently the minimization of $\mbox{tr}_1[h \g]$ on the `red' polytope will lead to a higher energy than the one on the `blue' polytope. This consideration confirms that for non-interacting fermions all the information about the excitation structure is contained in the boundary of the spectral polytopes. For weakly correlated systems, this is still approximately true since the minimizers will still lie very close to the boundary of the spectral polytopes (the total energy functional is in leading order indeed still linear because the pair interaction strength is assumed to be small). For stronger correlation this may change considerably. Yet, it is still true that by considering some $\wb^{(B)}$ sufficiently different to $\wb^{(A)}$ the corresponding polytope $\Sigma(\wb^{(B)})$ is getting so small that it does not contain the ground state vector $\bd{\lambda}^{(A)}$ anymore.

In summary, all these considerations suggest that the ground state functional may serve as a first approximation to $\,\Fbw$: At least for weakly interacting systems, the energy difference $E_{\wb} \leq E_{\wb'}$ (where $\wb' \prec \wb$) is primarily a direct \emph{geometrical} consequence, following from $\ebwp \subset \ebw$ and $\Sigma(\wb') \subset \Sigma(\wb)$, respectively. In particular, knowing the boundary of $\Sigma(\wb)$ is sufficient for determining approximately $E_{\wb}$.
A more detailed analytical investigation reveals that using the ground state functional in $\wb$-RDMFT as an approximation for $\Fbw$ would always yield in the minimization \eqref{Emin} a lower bound to $E_{\wb}$. Understanding for which systems this bound is sufficiently close to the exact value $E_{\wb}$ would require some extensive numerical testing.

\end{document}